# Fast Ensemble Smoothing

S. Ravela                    D. McLaughlin

**Ralph M. Parsons Laboratory**

**Department of Civil and Environmental Engineering**

**Massachusetts Institute of Technology**

**Cambridge, MA 02139**

1/21/06

**Corresponding Author:**

Sai Ravela, Bldg. 48-208, 15 Vassar Street, Cambridge, MA 02139

Tel: 617-253-1969, Email: ravela@mit.edu

**Abstract**

Smoothing is essential to many oceanographic, meteorological and hydrological applications. The interval smoothing problem updates all desired states within a time interval using all available observations. The fixed-lag smoothing problem updates only a fixed number of states prior to the observation at current time. The fixed-lag smoothing problem is, in general, thought to be computationally faster than a fixed-interval smoother, and can be an appropriate approximation for long interval-smoothing problems.

In this paper, we use an ensemble-based approach to fixed-interval and fixed-lag smoothing, and synthesize two algorithms. The first algorithm produces a *linear* time solution to the interval smoothing problem with a fixed factor, and the second one produces a fixed-lag solution that is *independent* of the lag length. Identical-twin experiments conducted with the Lorenz-95 model show that for lag lengths approximately equal to the error doubling time, or for long intervals the proposed methods can provide significant computational savings.

These results suggest that ensemble methods yield both fixed-interval and fixed-lag smoothing solutions that cost little additional effort over filtering and model propagation, in the sense that in practical ensemble application the additional increment is a small fraction of either filtering or model propagation costs. We also show that fixed-interval smoothing can perform as fast as fixed-lag smoothing and may be advantageous when memory is not an issue.

## 1. Introduction

Many earth science investigations are concerned with the reconstruction of past events, usually over an extended time interval using measurements available throughout the interval. Such retrospective data analyses can often be posed as smoothing problems, of which there are three well-recognized classes (Gelb, 1975). The first is *fixed point* smoothing, which requires the estimate of system state at only a single time $t$. The second is *fixed interval* smoothing, which requires estimates at multiple times distributed throughout an interval $[0,T]$. The interval length may be constant or may increase over time. The third is *fixed-lag* smoothing, which only requires estimates in a lag window $W$ prior to the most recent measurement. This is equivalent to considering only the measurements in a window $W$ after estimation time (see Figure 1). Fixed lag smoothing problems can serve as approximations to fixed interval problems or they may be useful in their own right.

To provide context, it is useful to note some typical earth science applications of smoothing. Fixed-interval smoothing is of considerable interest in oceanographic research projects such as Estimating the Circulation and Climate of the Ocean (ECCO), (Stammer et al., 2001, 2002). In these applications, spatially distributed states are estimated over decadal time-scales from a large number of observations. Oceanographic fixed-interval smoothing is typically posed as a batch least-squares parameter estimation problem and solved with representer (Bennett, 1992) or optimal control methods (Bryson and Ho, 1975; Wunsch, 1996). The smoothed estimates provide valuable information about ocean dynamics over a range of time scales. In meteorology, fixed-lag smoothing has been proposed and tested for the Goddard Earth Observing System (GEOS), both for retrospective analysis of climate and for improved short-term forecasting (Zhu et al., 2003). In hydrology, both batch and sequential approaches to fixed interval smoothing are being considered for the processing of radio-brightness and backscatter measurements from the Hydrosphere (HYDROS) satellite mission (Entekhabi et al., 2004). HYDROS products will use smoothing to estimate evaporative fluxes during inter-storm dry-down periods and to construct seasonal and annual water budgets over large spatial scales.

Smoothing problems may be solved either in a batch form, where all estimates are derived simultaneously from all measurements, or in a sequential form, where estimates are derived recursively through time. Variational algorithms are typically used to carry out batch processing over a smoothing interval of fixed length (Mernard and Daley, 1996), and sometimes this fixed length interval moves through time. In contrast to variational methods, sequential methods may offer advantages in some applications. For example, sequential methods generally provide information on estimation uncertainty; they can accommodate time-dependent input or model errors without any increase in computational effort.

Unfortunately, traditional sequential smoothing methods such as the Rauch-Tung-Striebel algorithm (Rauch, 1963) are limited to linear problems. Fixed-lag smoothers have been developed based on the Kalman and extended Kalman filters. They retain the benefits of sequential smoothing, and use adjoint methods to propagate information backward in time (Todling et al., 1998, Verlaan 1998, Cohn et al. 1994, Biswas and Mahalanobis, 1973). Recently proposed ensemble Kalman smoother and the fixed lag ensemble Kalman smoother (Evensen and van Leeuwen 2000; Evensen 2003) can handle nonlinear problems without requiring the derivation of an adjoint model, but they are still computationally demanding as a function of the interval or lag length.

In this paper, we introduce two ensemble smoothing algorithms that can provide efficient solutions to very large nonlinear problems. One is a fixed-interval smoother whose computational effort depends linearly on interval length, and the second is a fixed-lag smoother with computational effort that does not depend on lag length. These new algorithms make sequential ensemble smoothing an attractive and practical alternative to batch methods.

When smoothing is carried out with a sequential algorithm it is convenient to divide the process into three related steps. The first uses a dynamic model to move the system state forward through time (propagation), the second updates the current propagated state with current observations (filtering), and the third updates previous states with current observations (smoothing). In the spatially distributed estimation problems of most interest in earth science, the number of states to be estimated depends on the size of the computational grid used for the spatial discretization. The number of discrete times at which estimates are required in fixed-interval and fixed lag smoothing problems depends on the interval or lag lengths. Both the space complexity (the dependence of computational effort on state dimension) and time complexity (the dependence of computational effort on interval or lag length) of a smoothing algorithm have important impacts on computational feasibility. Both types of improvements are needed to obtain very efficient sequential smoothing algorithms for operational applications.

Various approximate strategies for improving the space complexity of large filtering and smoothing problems have been developed (Zhou et al., 2005; Whitaker and Hamill, 2001; Houtekamer, 2001; Hamill et al., 2001; Gaspari and Cohn, 1999). The methods described in this paper focus on improvements in the *time* complexity of fixed interval and fixed lag smoothers.

In Section 2 of this paper we review the theoretical framework for ensemble-based smoothing. We develop the fixed interval smoother in Section 3 and the fixed lag smoother in Section 4. In Section 5 we describe identical twin experimental assessments of both smoothers, using a Lorenz-95 dynamic model (Lorenz and Emanuel, 1998). We conclude in Section 6 with a discussion of computational issues relevant to smoothing.

## 2. Formulation of the Ensemble Smoothing Problem

The smoothing problem is typically defined in terms of a set of state and measurement equations. We suppose that these equations are provided in the following discretized form:

$$x_{t+\Delta t} = f(x_t, u_t) \tag{1}$$

$$y_{t'} = g(x_{t'}) + v_{t'} \tag{2}$$

In these equations, $x_t$ is a system state vector with an uncertain initial condition $x_0$, $u_t$ is a vector of uncertain model inputs (not necessarily additive), $y_{t'}$ is the measurement vector at measurement time $t'$, $v_{t'}$ is a vector of additive random measurement errors, and $\Delta t$ is the model time step. The random variables $x_0$ , $u_t$, and $v_{t'}$ are assumed to have known probability distributions and the measurement error vectors at different measurement times are assumed to be independent. The nonlinear functions $f$ (·) and $g$(·) represent spatially and temporally discretized models of the system dynamics and measurement process.

The objective of smoothing is to characterize the state $x_t$ at one or more estimation (or analysis) times using all measurements $y_{t'}$ taken at the *discrete* times $t'$ in the set $\tau$ [0,T] of measurement time in the interval [0, T]. The sequential ensemble smoothing algorithms we consider here are designed to estimate the conditional mean $E[x_t \mid y_{\tau[0,T]}]$, which is the minimum variance estimate of $x_t$, given $y_{\tau[0,T]}$ (Jazwinski, 1970). These algorithms rely on the fact that the conditional mean can be derived in closed form for the special case of jointly normal states and measurements. This closed form conditional mean solution can sometimes provide a useful approximation even when the requirements of joint normality are not met (for example, in the Lorenz 95 example considered later in this paper). It is particularly convenient because it relies on only the first two moments (means and covariances) of the states and measurements. The sequential ensemble smoothing approach proposed by Evensen and van Leeuwen (2000) derives these moments from random samples drawn from the input and measurement error probability distributions.

Evensen (2003, 2004) introduced a new interpretation of the ensemble Kalman filter and smoother that extends earlier work (Evensen and Van Leeuwen, 2000). This interpretation forms the basis for our algorithms and therefore, we start by discussing the filter, and then consider a simplified problem of updating an ensemble with observations at a future time.

Filtering can be thought of as updating a forecast ensemble with an observation at the current time. Let $A_t^f$ be an $n$ by $N$ ensemble matrix at time $t$, containing $N$ *forecast* state replicates of state dimension $n$ in each column. The updated estimate at $t$ can be written as (Evensen, 2003, 2004):

$$A_t^a = A_t^f X5_t \tag{3}$$

Here $A_t^a$ is an $n$ by $N$ ensemble matrix with each column a particular filter-analysis replicate. The form of the update in (3) states that the updated (analysis) ensemble is obtained by transforming the forecast state ensemble with the $N$ by $N$ matrix $X5_t$ constructed from the forecast ensemble and observations. Since $X5_t$ depends indirectly on $A_t^f$ the update expression in (3) is (weakly) nonlinear. The mean of the analysis ensemble is a sample estimate of $E[x_t | y_{0:t}]$.

It should be noted that the update formulation in (3) can be arrived by taking several different routes that consider robustness of the filter with respect to noise. In particular, these include Evensen (2004), the Ensemble Adjustment Kalman Filter (Andersen 2001, Tippett et al. 2003), and The Ensemble Transform Kalman Filter (Bishop et al., 2001). Shortly, we will see that the form of mixing an ensemble in (3) is the basis for the algorithmic synthesis of smoothing. Therefore, our algorithms are valid when any of these methods are used to generate the filter solution, provided the filter can be expressed in the same form as (3).

Filtering can happen only when there are observations. To make the subsequent discussion simpler, it is useful to view the matrix $X5_t$, defined in filtering (3), via the following transformation (Evensen, 2003), that makes it possible via a simple switch, to define a filtering operation at *every* time $t$:

$$X5_t = \begin{cases} I_N + X4_t & if \quad t \in \tau[0,T] \\ I_N & otherwise \end{cases} \qquad (4)$$

Here, $\tau$ is the set of measurement times, and $I_N$ is an $N \times N$ identity matrix and $X4_t$ is the transformation that produces an *analysis increment* to the forecast (see Evensen 2003). These equations define filtering everywhere on the interval; a filter analysis state is simply either a filtered ensemble (when observations are available) or forecast ensemble (when they aren't).

Now suppose we wish to update a prior estimate at some fixed *smoother-analysis* time $t$ with a single measurement at some future time $t'$ relative to it. In general, the smoother-analysis time, $t$, does not need to coincide with an observation time. A *smoothed* estimate for an ensemble using an observation at some future time $t'$ is shown by Evensen (2003), to be

$$A_t^s = A_t^a X5_{t'} \qquad (5)$$

It should be noted that $X5_{t'}$ is computed from the forecast ensemble and observations at time $t'$ (Evensen, 2004). This smoothed update is for a single smoothing-analysis time using a single measurement time.

## 3. Fixed Interval Smoothing

Fixed interval smoothing extends the single analysis time, single measurement time update of (5) to include multiple analysis and measurement times, all within a specified smoothing interval $[0, T]$. There are various ways to do this, some more efficient than others. The following subsections describe two options with differing time complexities.

### 3.1 The V1 algorithm

Evensen (2003) suggests a fixed interval smoother implementation that is a recursive application of (5). This update equation can be used to solve the fixed-interval problem if the interval is expanded sequentially, one measurement at a time. Suppose we have a filtered ensemble of state replicates at some measurement time $t''$ that lies in the set of measurement times $\tau(0, T)$ (see Figure 1). We use these replicates to initialize the state equation, which propagates the ensemble from $t''$ to a later measurement time $t'$ to yield a forecast $A_{t'}^{f}$ at $t'$. Then (3) is used to update $A_{t'}^{f}$ with the observations at $t'$, yielding a new filtered estimate $A_{t'}^{a}$ at $t'$. In addition, (5) is used repeatedly to smooth the most recent set of stored state replicates at all previous analysis times in the interval ($0 \leq t < t'$), one by one (this process can be parallelized) using the mixing matrix $X5_{t'}$. The filtered estimate at $t'$ provides the initial condition for a new forecast to the next measurement time. Then the entire process repeats, until $t' \geq T$.

Note that the state ensemble is initialized at $t = 0$ with replicates drawn from the initial condition probability distribution. Also, random input replicates drawn from the input probability distribution are included in the state equation during the propagation step. When $t' \geq T$ smoothed estimates have been computed for all analysis times in the fixed interval $[0, T)$ and a filtered estimate has been computed at the final time $t = T$. This smoothing algorithm, which we denote V1, consists of two nested loops that increment measurement times and analysis times.

A pseudocode implementation of V1 is provided in Appendix A. In this code, the user supplies the program with an ensemble of random replicates of the initial state, an observation operator, observation values, the model time step, the smoothing interval length, the set $T_o$ of model-time steps where observations are available, and a set $T_s$ of model-time steps where analyses are desired. This pseudocode provides the framework for a general smoothing program that can handle variable measurement and analysis times.

The computational cost of V1 can be readily evaluated in terms of the component costs of propagation, filtering, and smoothing. We define $C_m$ to be the computational cost of propagating the entire ensemble for one time step with the dynamic model (computation of $x_{t+1}$ from $x_t$, line13 of the pseudocode in Appendix A). The cost of a filter update at one measurement time $t'$ is $C_f$ (computation of $X5_{t'}$ and multiplication by $A_{t'}^f$, lines 4). The cost of updating one smoothing analysis time $t$ with one future observation $t'$ is $C_u$ (multiplication of $A_t^f$ by $X5_{t'}$, line 8).

The dependence of computation time on interval length is easiest to see if we consider a simplified scenario. Suppose that there are $M = T/\Delta t$ equally-spaced model time steps and $R$ observations spaced $r$ model steps apart in the fixed smoothing interval $[0,T]$. Also suppose there are $s$ analysis times between successive observations or $S = sR$ analysis times in the smoothing interval. The cost of V1 can be written as (for a list of symbols, see Appendix E):

$$C_{V1} = MC_m + \sum_{j=1}^{R}\left[C_f + \sum_{k=1}^{sj} C_u\right] \qquad (6)$$

$$= MC_m + RC_f + \sum_{j=1}^{R}\sum_{k=1}^{sj} C_u \qquad (7)$$

$$= MC_m + RC_f + sC_u R(R+1)/2 \qquad (8)$$

Since the total cost of filtering is $MC_m + RC_f$ the *incremental* cost of V1 fixed interval smoothing over filtering is:

$$\Delta C_{V1} = C_u sR(R+1)/2 = C_u s\, O(R^2) \qquad (9)$$

The quadratic dependence on the number of measurements (or interval length) obtained for this special case also applies for *unevenly* spaced analysis times. It is a direct result of the need to reevaluate (5) at all previous analysis times whenever a new measurement becomes available. Section 5 shows that quadratic cost dependence can make fixed interval smoothing with V1 quite expensive.

### 3.2 The FBF algorithm

We now describe a more efficient algorithm for the fixed interval smoother. In this approach, the entire interval smoothing operation is cast as a total of $R$ filter updates and $S$ smoothing updates, with no extra computation. The is done by our algorithm in three passes: 1) a forward (F) filtering pass, 2) a backward (B) pass that computes and stores products of X5 matrices, and 3) a second forward (F) pass that computes smoothing estimates at all analysis times. We call this three pass forward-backward-forward algorithm FBF.

The first forward pass of FBF computes filtered estimates from (3). States at times ($t < t'$) before the current measurement at $t'$ are not smoothed during this pass but ensemble matrices of filtered estimates are stored at all analysis times and X5 matrices are stored at all observation times. After the first forward pass of FBF we have an ensemble $A_t^f$ that is conditioned on all measurements taken through time $t$, at each $t$ in the fixed interval. Suppose that we want to update this ensemble with measurements obtained after $t$. This (smoothing) update may be written as (Evensen, 2003):

$$A_t^s = A_t^a \prod_{t' \in \tau_{(t,T]}} X5_{t'} = A_t^a X6_t \qquad (10)$$

Here $A_t^s$ is an $n$ by $N$ ensemble matrix of fixed point smoothing replicates evaluated at $t$. Each term in the product (10) incorporates the measurement at a particular future time, through a corresponding $X5$ matrix.

In the backward pass of FBF the $N$ by $N$ matrix is $X6_t$ computed, starting at $t = T$ and progressing backward to $t = 0$. This product can be computed sequentially and to see this it is conceptually useful to associate an $X5_t$ and $X6_t$ with each model time step and initialize $X5_t$ at unobserved model time-steps to be identity. Then, at any model time-step $t = k\Delta t$, the update

$$X6_t = X6_{k\Delta t} = \prod_{i=k+1}^{M} X5_{i\Delta t} \qquad (11)$$

may be written as a recursion

$$X6_{k\Delta t} = X5_{(k+1)\Delta t} X6_{(k+1)\Delta t} \qquad (12)$$

with $k\Delta t < T$, $X6_T = I_N$, and $X5_{j\Delta t} = I_N$ $\forall$ j$\Delta t \notin \tau_{[0,T]}$ as the conditions. Since $X5_{t'}$ at any unobserved model step t' is $I_N$, therefore in practice it need not be represented or multiplied. Therefore, at an analysis time $t$, $X6_t$ is simply the product of $X5_{t'}$ at all future observation times $t' > t$ in the interval. As an $X6_t$ is computed, any future $X5_{t'}$ still remaining are deleted, leading to (R+1) by N by N additional storage. The backward pass of FBF (12) accumulates all the information needed to apply (10), which is done in the final forward pass.

In the final forward pass, smoothed estimates at any given analysis time $t$ are obtained simply a multiplying $X6_t$ with the filtered estimates stored at $t$ (from the first forward pass), as indicated in (10). The mean of the updated ensemble obtained at the end of the final forward pass is a sample estimate of $E[x_t | y_{\tau[0,T]}]$.

Note that FBF carries out the filtering and smoothing operations in separate passes while V1 simultaneously filters and smooths states as it moves through the fixed interval in a single pass. The last forward pass has been presented here for convenience. The third pass, if used as is, can be implemented in parallel or when implemented sequentially can be combined with the backward pass in the following manner. Just after an $X6_t$ is computed in the backward pass, a smoothed analysis can be issued using the filtered ensemble stored from the first forward pass. However, the time complexity of this two-pass version is just the same as the three pass version. We will use the three-pass algorithm in experiments here, for convenience.

A pseudocode implementation of FBF is provided in Appendix B. Its input requirements are exactly the same as V1 and it incorporates all elements of the preceding discussion. The FBF pseudocode also is able to handle variable measurement and analysis times. The computational cost of the FBF algorithm can be derived in a manner similar to that used for V1. Now consider again the simplified scenario used in V1, where there are $R$ total observations on the interval, $M$ model integrations and there are $S=sR$ total smoothing analysis times. Using the same definitions of unit costs, we obtain:

$$C_{FBF} = MC_m + \sum_{j=1}^{R}\left[C_f + Cx\right] + \sum_{j=1}^{sR} C_u \qquad (13)$$

$$= MC_m + RC_f + RC_x + sRC_u \qquad (14)$$

Here $C_x$ is a new unit cost, defined as the additional cost of multiplying two matrices of dimensions $N$ by $N$, where $N$ is the ensemble size (line 15, Appendix B) . This is the cost of computing $X6$ for the current observation time from X5 and X6 in the immediate future. As before, the total cost of filtering only is $MC_m + RC_f$ so the incremental cost of FBF fixed interval smoothing over filtering is:

$$\Delta C_{FBF} = R\left[C_x + sC_u\right] \qquad (15)$$

The linear time complexity of FBF is a direct result of the computation and storage of X6 matrices on the backward pass. This intuitive idea makes it possible to update each filtered replicate only once, rather than repeatedly, as is done with V1. The FBF algorithm requires the additional storage of $R+1$ matrices, each of size $N$ by $N$. This is a small addition to the storage that both V1 and FBF need in order to save the $S$ filtered ensemble matrices, which are each $n$ by $N$, especially when $n >> N$.

The linear dependence of FBF on the number of measurements (or the interval length) is a major improvement over the quadratic dependence of V1. It is shown in Section 5 that this means ensemble smoothing is just as feasible as ensemble filtering for problems with large smoothing windows.

## 4. Fixed Lag Smoothing

Fixed lag smoothing updates states within a fixed time window prior to the current observation. If the observation is at $t$, then fixed lag smoothing implies updating states in [$t$ - $W$ $\Delta t$, $t$), where $W$ is the lag length. Equivalently, we can say that a state at $t'$ is updated with observations in the window ($t', t' + W\Delta t$). As estimation proceeds, the fixed lag window moves through the larger smoothing interval [0, $T$], where $T > W\Delta t$ can have a fixed value or can increase *indefinitely*. Each of these views of fixed-lag smoothing leads to different algorithms. The following subsections describe two fixed-lag smoothing options with different computational characteristics.

### 4.1 The V1-lag algorithm

Although the V1 algorithm is less efficient than FBF for fixed-interval smoothing problems, it offers some advantages for fixed-lag smoothing. The fixed-lag version of V1 follows from Evensen (2003). It is the same as the fixed interval version except that when a measurement is encountered at $t'$, (5) is used to update replicates only at analysis times in the lag window [$t'$ - $W$ $\Delta t, t'$), rather than at all previous analysis times. As in the fixed interval case, a filter update at $t'$ is used to initialize the propagation to the next measurement time. This smoothing algorithm, which we call V1-Lag, consists of two nested loops that increment measurement times and analysis times within the lag window.

A pseudocode implementation is provided in Appendix C. This code implements fixed lag smoothing over the interval $[0, T]$ and uses the inputs supplied as V1 and FBF. In addition, the user specifies the lag length $W$ in units of model time steps. The pseudocode presented in Appendix C is general in the sense that it handles non-uniform observation and analysis times. In addition, the pseudocode can be changed to handle an indefinitely growing interval.

The computational cost of V1-lag can be evaluated in much the same way as V1 and FBF. In order to facilitate comparison with the fixed interval smoothing algorithms we consider the case of using V1-Lag to find fixed lag smoothed estimates throughout $[0, T]$. As before, we assume there are $R$ equally-spaced observations with $r$ model time steps per observation and $s$ analyses per observation. Also, we assume that $L = W/r$, the number of observations in the window, is an integer. Then, with every new observation $sL$ prior states are updated. The computational cost is then:

$$C_{V1-lag} = MC_m + \sum_{j=1}^{R}\left[ C_f + \sum_{k=1}^{\min(sj,sL)} C_u \right] \tag{16}$$

$$= MC_m + RC_f + C_u \sum_{j=1}^{R} \sum_{k=1}^{\min(sj,sL)} 1 \tag{17}$$

$$= MC_m + RC_f + sC_u O(RL) \tag{18}$$

The incremental cost of V1-lag over filtering is therefore:

$$\Delta C_{V1-lag} = sC_u O(RL) = \frac{s}{r} C_u O(RW) \tag{19}$$

This algorithm has a linear, rather than quadratic, dependence on the lag-length and is efficient for small lags. As $W$ gets larger V1-lag approaches V1 and thus it is more expensive to implement V1-lag on the entire fixed smoothing interval than to use FBF. However, V1-lag has the advantage of requiring storage of $sL$ $n$ by $N$ ensemble matrices, whereas V1 and FBF require storage of $O(sR)$ matrices of the same size. We will examine these issues further later in this section.

## 4.2 The FIFO-lag algorithm

It is possible to develop a fixed lag algorithm that relies on some of the concepts introduced in the FBF fixed interval smoother. As the fixed lag window advances through time an ensemble of forecast estimates is computed, as in the ensemble Kalman filter or the first forward pass of FBF. However, the forecast replicates at analysis time $t$ are updated only with measurements in the window $(t, t+W\Delta t]$, rather than all future measurements, using a truncated version of (10)

$$A_t^s = A_t^a \prod_{t' \in \tau_{(t,t+W\Delta t]}} X5_{t'} = A_t^a X6_t \quad ; \quad t = [0,T] \tag{20}$$

If the final smoothing time $T$ is constrained to a fixed value, this moving update needs to be supplemented with the following terminal conditions:

$$X5_{t'} = I_N \quad ; \quad t' > T \tag{21}$$

and $I_N$ is an $N$ dimensional identity matrix. In the fixed lag smoothing problem $X6_t$ and $A_t^s$ do not need to be computed on separate passes. Instead, $X6_t$ can be computed with a forward recursion. To see this, initialize X5 at all unobserved model steps:

$$X5_t = I_N \, ; \, t \notin \tau[0, T + W\Delta t] \tag{22}$$

And initialize X6 at $t = 0$:

$$X6_0 = \prod_{t' \in \tau(0, W\Delta t]} X5_{t'} \, . \tag{23}$$

Then the following recursion defines fixed lag smoothing:

$$X6_t = X6_{k\Delta t} \tag{24}$$

$$= \prod_{i=k+1}^{k+W} X5_{i\Delta t} \tag{25}$$

$$= X5_{k\Delta t}^{-1} X6_{(k-1)\Delta t} X5_{(k+W)\Delta t} \tag{26}$$

$$X6_t = X5_t^{-1} X6_{t-\Delta t} X5_{t+W\Delta t} \tag{27}$$

This recursion builds up the initial $X6$ value in (23) by multiplying $X5$ matrices in the window $(0, W\Delta t]$. Once this is done, subsequent $X6$ values are obtained by pre- and post–multiplying X6 with $N$ by $N$ matrices at every model step, as indicated in (26) or (27). The matrix X5 is invertible so long as the replicates in the state ensemble are linearly independent, which they will be for any state equation of practical interest. The entire fixed lag smoothing procedure is therefore done on a single forward pass.

We call this algorithm FIFO-lag, because it implements fixed-lag smoothing via a *first-in-first-out queue*. New information is added to the front of the queue and old information is removed from the back of the queue. In practice, matrix multiplications in (26) can be omitted when either or both of the X5 matrices are identities. This reduces computational effort when the measurements are sparse in time. A pseudocode implementation of FIFO-lag is provided in Appendix D. This code uses the same inputs supplied as V1-lag. It handles nonuniform observation and analysis times and can be changed to accommodate an indefinitely growing interval.

$$C_{FIFO-lag} = MC_m + RC_f + RC_{X6} + sRC_u \tag{28}$$

Here, $C_{x6}$ is the cost of implementing the recursion in Equation (26). The incremental cost of FIFO-lag over filtering is therefore:

$$\Delta C_{FIFO-lag} = RC_{X6} + sRC_u = R[C_{X6} + sC_u] \tag{29}$$

This cost is independent of the lag length *W*. The FIFO-lag algorithm is somewhat more expensive than *FBF* but is less costly than V1-lag after a certain lag-length. We will examine this tradeoff further in Section 5. In terms of memory, FIFO-lag only requires storage of ensemble members and *X5* matrices at observation times within the lag window X5 matrices at other times are identities and need not be stored. This makes FIFO-lag especially attractive when storage is limited.

We now turn to the computational performance of the four smoothing algorithms. The unit costs defined earlier can be written as $C_u = nN^2$, $C_x = N^3$ and $C_{x6} \sim 3N^3$. There, $n$ is the state size and $N$ is the ensemble size (see Appendix E for a list of symbols). The incremental costs of the four algorithms are then:

$$\Delta C_{V1} = nN^2 sR(R+1)/2 \qquad\qquad \Delta C_{FBF} = R(N^3 + snN^2)$$

$$\Delta C_{V1\text{-lag}} = LR\,sn\,N^2 \qquad\qquad \Delta C_{FIFO\text{-lag}} = R(3N^3 + snN^2)$$

Please recall that $R$ is the number of observations over the inverval, L is the lag-length in number of observations, and $s$ is the ration of number of smoothing analysis times to number of observations on the interval. From these numbers, we may conclude the following:

1. **Fixed Interval Smoothing: FBF is faster than V1.** A comparison of the costs of the $\Delta C_{V1}$ and $\Delta C_{FBF}$ fixed interval smoothing algorithms suggests that V1 will require more computational time when $R > 1 + \dfrac{2N}{sn}$. This is certainly the case for large problems, where the ratio of replicates to states ($N/sn$) is typically small and the number of measurement times in the fixed smoothing interval ($R$) is large. The superior performance of FBF is expected since the V1 computation time grows quadratically with the fixed interval length while the FBF computation time grows linearly.

2. **Interval vs. Lag: FBF is generally faster than V1-lag but it requires more memory.** A comparison of the costs of the $\Delta C_{V1\text{-}lag}$ and $\Delta C_{FBF}$ smoothing algorithms suggests that V1-lag will require more time to compute smoothed estimates throughout a fixed interval if

   $L > 1 + \dfrac{N}{sn}$. In practice, when $N/sn < 1$, this leads to the interesting conclusion that FBF is faster for comparable accuracy than V1-lag, after a very short lag-length, even though it processes all measurements rather than just those in the lag window. This implies FBF is faster for most practical fixed lag smoothing problems. Computationally, the only reason to prefer the fixed-lag approach is the substantial memory savings it provides. However, when the system is nonlinear, it may be the case that fixed-interval smoothing, over a long interval, is not the best solution to the smoothing problem, but rather a fixed-lag solution is. In these cases too, fixed-lag smoothing will be preferable.

3. **Interval vs. Lag: FBF is slightly faster than FIFO-lag but it requires more memory.** A comparison of the costs of $\Delta C_{FIFO\text{-}lag}$ and $\Delta C_{FBF}$ suggests that FIFO-lag is more expensive by a *fixed* factor. Essentially the difference between FBF and FIFO-lag is one extra multiplication between two $N$ by $N$ matrices and one inversion of an $N$ by $N$ matrix, which as a ratio between $C_{FIFO\text{-}lag}$ and $C_{FBF}$ is a fixed number. Since FIFO-lag consumes the same memory as V1-lag, it shares this advantage over FBF.

4. **Fixed Lag Smoothing: FIFO-lag is faster than V1-lag, beyond a certain lag length.** Finally, a comparison between the costs of FIFO-lag and V1-lag suggests that V1-lag is more expensive when $L > 1 + \dfrac{3N}{sn}$. This result is independent of the interval length. Therefore, when longer lag lengths are needed to approximate the fixed interval smoothing solution it is preferable to use FIFO-lag.

It should be emphasized here that in most practical applications $N/n << 1$, and then the benefits of the proposed algorithms are computationally even better. In the case that the number of smoothed analysis times is less than the number of observed time i.e. s<1, these bounds can worsen, but not in most practical situations where the ratio N/sn remains much smaller than 1.

## 5. Numerical Comparison of the Smoothing Algorithms

We now compare the various fixed interval and fixed lag smoothing algorithms described in Sections 3 and 4, using an identical-twin experiment based on the a Lorenz-95 system (Lorenz and Emanuel, 1998). The continuous time Lorenz equations are:

$$\frac{dx}{dt} = -x_{i-2}x_{i-1} + x_{i-1}x_{i+1} - x_i + u$$

where $i = 1, \dots n$ is cyclical (i.e. $x_0 = x_n$, $x_{-1} = x_{n-1}$, $x_{n+1} = x_1$) and can be interpreted as a surrogate spatial index. The constant forcing term is $u = 8$ in all simulations. This model is integrated forward using a fourth-order Runge-Kutta scheme (Press et al., 1988). The integrated Lorenz-95 equation system defines the discrete time propagation function $f(\cdot)$ in (1). The Lorenz-95 model is an interesting nonlinear (chaotic) system that is convenient to use in identical-twin experiments with moderate state sizes. It has some relevance to meteorological problems because it simulates circulation on a latitude circle (Lorenz and Emanuel, 1998).

Our identical-twin experiments generate a true state vector of dimension $n = 100$ from a random zero-mean Gaussian initial condition with a standard deviation of 2.0. The equation system is integrated from this initial condition for 8192 steps to remove transients. This integrated value defines the true system state at the beginning of the smoothing interval ($t = 0$). The system is integrated further until $t = T$. Synthetic observations are generated at specified measurement times by adding uncorrelated zero-mean Gaussian noise with a standard deviation of 0.2 to the true state. Since all states are measured directly, our experiment the measurement operator $g(\cdot)$ in (2) is just a binary incidence matrix.

A first guess of the true state at $t = 0$ is obtained by perturbing the true initial state by a vector of uncorrelated zero-mean Gaussian random variables with standard deviation 2.0. Then an ensemble of 100 random initial condition replicates is obtained by perturbing the first guess with 100 vectors of uncorrelated zero-mean Gaussian random variables with a standard deviation of 1.0. The ensemble smoothing algorithms compared in our experiments generate estimates at every time step. The model time step $\Delta t$, the fixed interval length $T$, and the measurement set $\tau$ $(0, T)$ can all be varied. All calculations are carried out on a 2.8GHz Intel processor with 512KB cache and 2GB memory. The code was implemented in MATLAB running under Linux with a SCSI disk (LSI53C010) and a Fusion MPT board with a normal EXT-3 file system.

In our first experiment the system is integrated in the interval [0, 1] with a dimensionless time step Δt =0.01, giving $M = 100$. The constant observation interval is 0.05, giving $r = 5$. The state is observed at every other location so $m = n/2 = 50$. Smoothed analyses are produced from smoothed ensemble estimates at every model time step, therefore $S = 100$ (the last time-step cannot be smoothed), and $T = [0, 1, 2,…, 99]$. Lorenz and Emanuel (1995) associate the dimensionless time step 0.01 with a real time of 1.2 hrs, implying that our measurement interval corresponds to 6 hrs. and our fixed interval smoothing window is 120 hrs..

Figure 2 compares the estimation error (over all states) obtained from an ensemble Kalman filter with those obtained from the V1 and FBF ensemble smoothing algorithms. The error in each state is the difference between the estimate (smoothed analysis ensemble mean) and the known true values. The V1 and FBF smoothing algorithms give the same errors, which are smaller than the ensemble filter errors at all times except the end points. This reflects the fact that the smoothers use all measurements at each time while the filter uses only measurements from earlier times.

Our second experiment uses the same inputs as the first but considers the two fixed lag smoothers. Figure 3 compares the root-mean-squared errors obtained from V1-lag and FIFO-lag for fixed lag lengths of $L =1, 5, 9$, and 13 measurements, corresponding to $W = 5, 25, 45$ and 65 model time steps. In every case the FIFO-lag and V1-lag estimates and errors are the same. When $L = 1$ the fixed lag smoothing error is nearly the same as for the ensemble Kalman filter. As the lag length increases the fixed lag smoother errors decrease, approaching those of the fixed interval smoother. The fixed lag smoothers give essentially the same result as the fixed interval alternative when $L$ is greater than 9. The Lorenz-95 model has an error doubling time of 2.1 days (0.45 in dimensionless time), which corresponds to $L = 9$. This suggests that fixed lag smoothers can give very good approximations to fixed interval smoothers if the lag is of the order of the system's memory.

We now turn to the computational performance of the four smoothing algorithms. In the following experiments these algorithms are compared by varying the interval and lag lengths, while the state size and number of replicates remain fixed at $n = 100$ and $N = 100$, respectively. States are completely observed at every time step so $m = n$ and $r = 1$. Analyses are also produced at every model time step, so $s = 1$. The results presented here verify prior hypotheses:

1. **Fixed Interval Smoothing: FBF is faster than V1.** Figure 4 indicates how the added work of smoothing compares to model propagation only and to filtering only. It is apparent that V1 smoothing takes much more time than filtering, becoming prohibitively expensive for large problems. By contrast, FBF smoothing adds only a modest amount of computational effort to filtering, with the computational time growing with interval length at approximately the same rate as filtering alone.

2. **Interval vs. Lag: FBF is generally faster than V1-lag but it requires more memory.** In our experiments, where $N = n = 100$ and $s = 1$, V1-lag is more expensive than FBF when the lag $L > 2$. We have seen that lags of 9 or more are needed in our Lorenz-95 example in order for fixed lag estimates to be as accurate as the comparable fixed interval estimates. Thus, when the lag-length is long, FBF may be preferable. Figure 5 illustrates these conclusions with a comparison of the FBF and V1-lag computational times required to obtain smoothed estimates throughout a fixed interval of lengths 100, 500, and 900. It is apparent that FBF requires less computation time than V1-lag for lags greater than 2.

3. **Interval vs. Lag: FBF is slightly faster than FIFO-lag but it requires more memory.** Figure 6 depicts the computation time as a function of lag-length for a smoothing problem that extends through the entire interval $[0, T]$, where $T = 100$, 500 and 900. All inputs are the same as described above. The figure indicates that FIFO-lag costs about 1.45 times the FBF cost. Since FIFO-lag consumes the same memory as V1-lag, it shares this advantage over FBF.

4. **Fixed Lag Smoothing: FIFO-lag is faster than V1-lag, beyond a certain lag length.** In our conservative example with $s = 1$ and $n = N$, we predict the threshold lag to be $L = 4$. Indeed, a comparison of Figure 5 and Figure 6 indicates that the computational experiment is in excellent agreement with this prediction. Therefore, when longer lag lengths are needed to approximate the fixed interval smoothing solution it is preferable to use FIFO-lag.

## 6. Parallel Extensions

In most practical applications, it would be reasonable to assume the availability of parallelism, most often through a computational cluster. Here, we show that for *fixed-interval* applications, parallelism can *amplify* the advantages of the proposed algorithm.

The V1 smoothing algorithm parallelizes easily. Every time a new X5 matrix is computed all the previous states are updated in parallel. If we assume the interval is *fixed*, then we have to update with $R$ *X5* matrices, each updating several prior ensembles. Since filtering and smoothing are not separated in V1, therefore, these $R$ updates proceed sequentially on the interval. In practice, if we suppose that there are $P$ parallel processors and ignore the communication cost of transporting *X5* matrices the complexity of V1 can be computed as: $\Delta C_{V1} \approx C_u \max(\frac{sR}{P}, 1) R$. In the *asymptotic* limit, when the number of parallel processors is more than $R$, then this cost reduces to

$$\Delta C_{V1} \approx C_u R \qquad (30)$$

Thus a parallelized version of V1 will be a linear time algorithm.

The FBF algorithm is also parallelizable. In the third pass, the update of each ensemble member with the corresponding *X6* matrix can be done in parallel. Thus the cost of FBF, with this level of parallelism, is

$$\Delta C_{FBF} \approx RC_x + \max(\frac{sR}{P}, 1) C_u$$

Again, taking the *asymptotic* limit, we obtain

$$\Delta C_{FBF} = RC_x + C_u \qquad\qquad\qquad (31)$$

It is clear from equations (30) and (31) that the incremental cost of FBF will be smaller. This is because the cost $C_x$ is $O(N^3)$ and the cost of $C_u$ is $O(nN^2)$, implying that when n>>N, as is the case in most practical applications $\Delta C_{FBF} < \Delta C_{V1}$.

The key distinction between the two methods is that even in the parallelized version, V1 has to repeatedly update states, sequentially, as new observations are processed. X5 matrices from a new observation cannot be incorporated until the previous ones have. In the parallelized FBF algorithm, X6 matrices are computed sequentially, and then the final smoothed updates are produced modulo as many parallel processors are available. Doing so is cheaper because it is cheaper to produce one *X6*, recursively, than it is to repeatedly update the ensemble.

It can be argued that if a sufficient number of parallel processors are available and information transfer across them is inexpensive, then V1-lag is also essentially independent of the lag length because updates to prior states can be conducted in parallel when a new X5 is synthesized. Nonetheless, in much the same way as V1, fixed lag smoothing must proceed sequentially as new observations are processed over the interval to generate the filter solution and X5s. The computational cost of V1-lag with a parallel scheme, where all smoother-analysis times are updated in parallel leads to the following computational cost:

$$\Delta C_{V1-lag} = \max(\frac{sW}{rP},1)RC_u$$

If there are sufficient number of parallel processors as the number of updates within a lag

window, then $\Delta C_{V1-lag} = RC_u$. Ironically, V1-lag cannot make use of any extra processing power

beyond the number of smooth-analysis times that lie in the lag window. Thus, for a lag-length of

1,2,…9 units, the extra processors, which will typically be much larger, will go unused.

There are number of ways in which either FBF or FIFO-lag can be adapted for parallel

fixed-lag smoothing. Here, we present the situation for FIFO-lag as an approximation to long-

term retrospective reanalysis, that is, as an approximation to the fixed-interval smoothing

problem. Parallelization can be brought into play in the FIFO-lag algorithm by splitting it up into

two passes. In the first, pass, the filter solution and X6s are computed. In the second pass, much

like FBF, the updates of states at smoother-analysis times are conducted in parallel. This has

direct advantage when *FIFO-lag is used to solve a fixed-interval smoothing problem*. The

incremental cost under this parallelization scheme becomes:

$$\Delta C_{FIFO-lag} = RC_{X6} + \max(\frac{sR}{P},1)C_u$$

As the number of parallel processors grows, so does the advantage of the parallelized version of FIFO-lag. In the asymptotic limit of sufficient number of parallel processors over the interval, we have: $\Delta C_{FIFO-lag} = RC_{X6} + C_u$ Since the state size n>> N typically, the ensemble size, therefore, for fixed interval approximated by fixed-lag, FIFO-lag quickly outperforms V1-lag. In fact, this advantage is even more pronounced when the lag window is small. V1-lag simply cannot use the extra computational power.

## 7. Conclusions

This FBF and FIFO-lag algorithms introduced in this paper are significant computational improvements to previous ensemble smoothing algorithms. These improvements yield a fixed interval smoothing solution that is linear in time and a fixed lag solution that is independent of the lag length. The FBF fixed interval smoother is faster than the V1 fixed interval solution suggested by Evensen for interval smoothing. FBF is faster than V1-lag, the lagged version of Evensen's smoother, past a small lag window length, but it requires more memory. FBF is faster but requires more storage than FIFO-lag. FIFO-lag is faster than V1-lag for moderate lags.

We believe that the FBF algorithm will yield efficient solutions for long fixed interval problems. FBF and FIFO-lag are both promising candidates for fixed-lag smoothing problems. When memory is a limitation, FIFO-lag is the best solution for the fixed lag problem or for the fixed interval problem if the lag length is longer than the system's memory. The FBF and FIFO-lag algorithms documented in the pseudocodes of Appendixes A through D can both be used with unequally spaced measurements and analysis times. The computational cost analyses presented here use equally-spaced measurements and analysis times only because this makes the dependence of cost on interval or lag length more apparent.

The fast ensemble smoothing algorithms presented here help make ensemble smoothing a practical option for retrospective analyses of large earth science data sets. Indeed, our work (the FIFO-lag algorithm) is now being used by other groups (Heemink et al., 2005) . We have not yet explored how the FIFO-lag algorithm can be adjusted to incorporate time-dependence and field-dependence of the optimal lag lengths over the interval.

The improvements in temporal computational complexity described here need to be combined with improvements in spatial computational complexity. Taken together, these improvements could greatly reduce the burden of measurement updating, leading to a situation where the dynamic model run time (and storage) rather than the Bayesian update step will be the factors that limit the application of ensemble smoothing concepts to practical problems.

**Appendix A: V1 -- Fixed interval smoothing algorithm**

A nominal implementation of ensemble smoothing suggested by Evensen (2003) is presented.

This algorithm is called *V1* in the text.

1: i ← 0

2: **while** $i \leq M$ **do**                 %% Loop over interval [0, MΔt]

3:    **if** $i \in T_o$ **then**             %% if $i\Delta t$ is an observation

4:       compute X5;    A ← A *X5           %% Filter the forecast.

5:         **for** *j = 0 to i-1* **do**       %% and loop through all previous states

6:            **if** $j \in T_s$ **then**         %% and update every smooth-analysis state $j\Delta t$.

7:                                   %% with X5 computed at $i\Delta t$

8:                A←Load('A',*j*); A ← A * X5; Save('A',j,A)

9:            **end if**

10:         **end for**

11:    **end if**

12: **if** $i \in T_s$ **then** Save('A',i,*A*) **end if**    %% Save filtered ensemble if it is a smoothing time

13: A← M(A, Δt); *i* ←*i + 1*            %% Integrate forward to next time step.

14: **end while**

The routine Load('A',j) loads the file with name A annotated with the value j. Thus for example, Load('A',5), loads 'A-5', which would be the filtered ensemble at model step 5. Similarly, Save('A',5,A) saves the matrix A in the file 'A-5'.

**Appendix B: FBF -- Fixed interval smoothing algorithm**

An improved algorithm, called FBF, for the interval smoothing problem using an ensemble framework. For convenience, each pass is printed on a separate page.

1: i$\leftarrow$0; k$\leftarrow$0
2: **while** $i \leq M$ **do**                    **%% PASS 1: Loop over interval [0,T]**
3:        **if** $i \in T_o$ **then**                    **%%** An observation is encountered
4:              Compute X5;  A$\leftarrow$ A *X5;        %% This is the filtering step
5:              k$\leftarrow$k+1; Save('X5',k,X5)       %% Save X5$_k$ for pass 2
6:        **end if**
7:        **if** $i \in T_s$ **then**                   %% If i$\Delta$t  is smoothed analysis time
8:              Save('E',i,[k+1,A]);                    %% Save filtered or forecast ensemble,
9:        **end if**                                    %% and an index for the next obs. (k+1)
10:        A$\leftarrow$ M(A, $\Delta$t); $i \leftarrow i + 1$        %% Step it: Integrate entire ensemble forward one step.
11: **end while**                        **%% First pass finished.**
                                **%% Start PASS 2**
12: X6$\leftarrow$load('X5',k); Save('X6',k,X6);    %% Initialize X6$_k$
13: **for** j = k - 1 to 1 **do**                    %% Loop over remaining observations.
14:        X5$\leftarrow$load('X5',j);
15:        X6$\leftarrow$ X5*X6;                       %% Back-multiply a cumulative product.
16:        Delete('X5',j);                            %% Then there is no need for keeping X5.
17:        Save('X6',j,X6);                           %% But save the accumulated product.
18: **end for**                        **%% End pass 2**
19: **for** $i \in T_s$ **do**                        **%% PASS 3: Loop over analysis times**
20:        [j, A] $\leftarrow$load('E',i);             %% Load filtered ensemble and X6 index.
21:        **if** j $\leq$ k **then**        %% Don't update ensemble starting from last observation.
22:              X6$\leftarrow$load('X6',j)            %% Now smooth.
23:              A $\leftarrow$ A * X6; Save('A',i,A)    %%  And save smoothed ensemble
24:        **end if**
25: **end for**                        **%% End PASS 3.**

**Appendix C: V1-Lag -- Fixed lag smoothing algorithm**

This pseudo code presents a nominal implementation of fixed-lag ensemble smoothing suggested by Evensen [2]. This algorithm is called *V1-lag* in the text.

```
1: i ← 0
2: while i ≤ M do                                   %% Loop over interval [0,T]
3:    if i ∈ T_o then
4:       compute X5;   A ← A *X5                     %% This is the filtering step
5:          for j = max(0, i-W) to i-1 do            %% Loop through at most W prior model steps
6:             if j ∈ T_s then                       %% Updating the states to be analyzed.
7:                A ← Load('A',j);
8:                A ← A * X5;
9:                Save('A',j,A)
10:            end if
11:         end for
12:   end if
13: if i ∈ T_s then Save('A',i,A) end if             %% Save filtered ensemble at smoothing time
14: A ← M(A, Δt); i ← i + 1                           %% Integrate forward.
15: end while
```

**Appendix D: FIFO-Lag -- Fixed lag smoothing algorithm**

1: i←0; X6 ←**I$_N$**
2: **while** $i \leq M$ **do**                                           %%% Loop over interval [0,T]
3:    X5 ← []                                          %%% Initialize X5 to be null matrix.
4: **if** $i \in T_o$ **then**                                      %%% An observation is encountered
5:    Compute X5;  A ← A *X5;                   %%% Filter and compute X5.
6: **end if**
7**:** Save('X5',i,X5);                                 %%% Save X5 for later fixed lag access.
8: **if** $i \in T_s$ **then** Save('A',i,A); **end if**          %%% Save ensemble at smoothing time
9: **if** i< W **then**                                     %%% If Lag window has not spun-up
10:    X6 ←MultR(X6,X5);                       %%% build up X6
11: **else**                                              %%% otherwise apply FIFO policy
12:    X$_w$←load('X5',i-W);                       %%% Load X5 at start of lag window
13:    X6 ← Mult(X$_w$,X6,X5);              %%% Pop out X5$_{i-w}$ from X6 and pop in X5$_i$ at end
14:    Delete('X5',i-W);                       %%% Delete X5 past lag window.
15:    **if** i-W $\in T_s$ **then**                       %%% update lagged state
16:       A$_w$←Load('A',i-W);  A$_w$ ← A$_w$*X6; Save('A',i-W,A$_w$);
17:    **end if**
18: **end if**
19: A← M(A, Δt); i ← i + 1                    %%% Integrate forward to next time step.
20: **end while**
21: **for** j = M-W+1 to M-1                          %%% Spin down lag window to all but last state
22:    X$_w$ ←Load('X5',j);
23:    X6 ← MultL(X$_w$,X6); Delete('X5',j);
24:    **if** $j \in T_s$ **then** A$_w$←Load('A',j); A$_w$ ← A$_w$*X6; Save('A',j,A$_w$); **endif**
25: **end for**

**function C** = MultR(A,B)
    **if ~isempty**(B)  **then** C = A*B; **else** C=A; **endif**
**endfunc**

**function C** = MultL(A,B)
    **if ~isempty***(A)* **then** *C = pinv(A)*B;* **else** *C = B;* **endif**  *%%*pinv is same as inverse.
**endfunc**

**function D** = Mult(A,B,C)
    D = MultL(A, MultR(B,C))
**endfunc**

**Appendix E: Symbols and their meanings**

$A_t^f$, $A_t^a$, $A_t^s$ : The forecast, filtered and smoothed ensembles at time $t$, respectively.

$C_f$: Cost of filtering per observation time.

$C_m$: Cost of propagating model through one model step.

$C_u$: The cost of updating an ensemble: multiplying an nxN matrix with an NxN matrix.

$C_x$: The cost of multiplying two matrices of size *NxN*.

$C_{x6}$: The cost of inverting an *NxN* matrix plus multiplying three *NxN* matrices.

$f$:    Model, system propagator.

$g$:    Observation operator.

$L$:    Lag-window in number of observations.

$M$:    Number of model steps in the interval.

$n$:    State size.

$N$:    Ensemble size.

$P$:    Number of parallel processors.

$r$:    Number of model steps per observation (uniform).

$R$:    Number of Observations.

$s$:    Number of smoother analyses per observation (uniform).

$S$:    Number of smoother analyses.

$t,t',t''$:    *T*ime indices on the interval.

$\Delta t$:    Model step size.

$\tau$:    The set of observed times in the interval.

$u_t$:    Forcing

$T_o$:    The set of observed times in model steps in the interval.

$T_s$:    The set of smoother analysis times in model steps in the interval.

$v_{t'}$:    Noise.

$W$:    Lag-window in number of model steps.

**References**

Anderson, J. L., 2001: An Ensemble Adjustment Kalman Filter for Data Assimilation. *Mon. Wea. Rev.*, 129, 2884-2903.

Bennett. A.F., 1992: *Inverse Methods in Physical Oceanography.* Cambridge Monographs on Mechanics and Applied Mathematics, Cambridge University Press.

Bishop, C. H., B. J. Etherton, and S. J. Majumdar, 2001: Adaptive Sampling with Ensemble Transform Kalman Filter. Part I: Theoretical Aspects. *Mon. Wea. Rev.*, 129, 420-436.

Bryson Jr., A. E., and Ho, Y. C., 1969: *Applied Optimal Control.* Blaisdell, Waltham, MA, USA.

Cohn, S. E., Sivakumaran N. S., and Todling R., 1994: A Fixed-Lag Kalman Smoother for Retrospective Data Assimilation, *Monthly Weather Review*, **122**, pp. 2838–2867.

Entekhabi, D., E. Njoku, P. Houser, M. Spencer, T. Doiron, Y. Kim, J. Smith, R. Girard, S. Belair, W. Crow, T. Jackson, Y. Kerr, J. Kimball, R. Koster, K. McDonald, P. O'Neill, T. Pultz, S. Running, J. Shi, E. Wood, and J. van Zyl, 2004: The Hydrosphere State (Hydros) mission: An Earth system pathfinder for global mapping of soil moisture and land freeze/thaw, *IEEE Trans. Geosci. Rem. Sens.,***42**, 2184-2195.

Evensen, G. and Van Leeuwen P. J., 2000: An ensemble Kalman smoother for nonlinear dynamics, *Monthly Weather Review*, **128,** 1852-1867.

Evensen, G., 2003: The ensemble Kalman filter: theoretical formulation and practical implementation. *Ocean Dynamics*, **53**, 343-367.

Geir Evensen, 2004: Sampling strategies and square root analysis schemes for the EnKF *Ocean Dynamics*, 54, 539-560.

Gaspari, G., and Cohn S. E., 1999: Construction of correlation functions in two and three dimensions. *Quarterly Journal of Royal Meteorological Society., ***125**, 723–757

Hamill, T. M., Whitaker J. S., and Snyder C., 2001: Distance-dependent filtering of background error covariance estimates in an ensemble Kalman filter. *Mon. Wea. Rev., ***129**, 2776–2790

Heemink, A. and Girardeau, P. Personal Communication, July 2005.

Houtekamer, P. L., 2001: A sequential ensemble Kalman filter for atmospheric data assimilation. *Mon. Wea. Rev., ***129**, 123–137

Jazwinski, A. H., 1970: *Stochastic Processes and Filtering Theory.* Academic Press, 376 pp.

Lorenz, E. and K. Emanuel, 1998: Optimal sites for supplementary weather observations: Simulation with a small model. *Journal of Atmospheric Sciences*, **55**, 399-414.

Me´nard, R., and R. Daley, 1996: The application of Kalman smoother theory to the estimation of 4DVAR error statistics.Tellus, **48A**, 221–237.

Press, W. H., B. P. Flannery, S. A. Teukolsky, and W. T. Vetterling. *Numerical Recipes in C, The Art of Scientific Computing*. Cambridge University Press, 1988.

H. E. Rauch, 1963 Solutions to the linear smoothing problem, *IEEE Transactions on Automatic Control*, vol. **8**, pp. 371-372.

Stammer, D., R. Bleck, C. Böning, P. DeMey, H. Hurlburt, I. Fukumori, C. LeProvost, R. Tokmakian, Wenzel, 2001: Global ocean modeling and state estimation in support of climate research. In: *Observing the Ocean in the 21st Century,* C.J. Koblinsky and N.R. Smith (eds), Bureau of Meteorology, Melbourne, Australia, 511--528.

Stammer, D., C. Wunsch, I. Fukumori, and J. Marshall, 2002: State estimation in modern oceanographic research, *EOS*, *Transactions, American Geophysical Union*, **83** (27), 289&294-295.

Tippett, M. K., J. L. Anderson, C. H. Bishop, T. M. Hamill, and J. S. Whitaker, 2003: Ensemble Square Root Filters. *Mon. Wea. Rev.*, 131, 1485-1490.

Todling R., Cohen, S. E., and Sivakumaran N. S., 1998: Suboptimal schemes for retrospective data assimilation based on the Fixed-lag Kalman smoother. *Monthly Weather Review*, **126**, 247–259.

Verlaan, M., 1998: Effcient Kalman Filtering algorithms for hydrodynamic models. Ph.D. thesis, Technische Universiteit Delft, 201 pp.

Whitaker, J. S. and Hamill, T. M., 2001: Ensemble Data Assimilation without Perturbed Observations, *Monthly Weather Review,* **130**, pp. 1913–1924.

Wunsch, C., 1996. The *Ocean Circulation Inverse Problem*, Cambridge University Press, 456 pp.

Zhu Y., Todling R., Guo J., Cohn S. E., Navon M., and Yang Y, The GEOS-3 Retrospective Data Assimilation System: The 6-Hour Lag Case, *Monthly Weather Review,* **131**, 2129-2150.

Zhou Y., McLaughlin, D., and Entakhabi, D., An Ensemble Multiscale Filter for Large Scale Land Surface Data Assimilation, 2005: submitted

**List of Figures**

1. Illustration of smoothing problems. Subfigure (A) depicts fixed-interval smoothing and subfigure (B) depicts fixed-lag smoothing. In subfigure (A), there are M=24 model integrations, with R=6 observations at r=4 equally spaced model steps 4,8,12,16,20 and 24. This figure is a graphical model for interval smoothing. It shows that the ensemble is integrated forward starting at time-step 0 (see "Integration"). When an observation is encountered (see row "Observations"), filtering is performed. For example, at $Y_4$ the corresponding $X5_4$ is generated and used to update the forecast ensemble ( $A_4^f$ ). It is also used to update forecast (or filtered) ensembles at all previous times where smoothed analyses are desired. In this example smoothed analyses are desired at time steps 0, 5, 10, 15 and 20 and may be between or at measurement times. Using the depicted scheme, it can be seen that the smoothed analysis at time 0 utilizes all 6 future observations, the smoothed analysis at time step 10 uses four future observations (filtering already incorporates all observations to time step 10), and so on. Equivalently, the observation at time step 4 influences all states at time steps 0-4 and, progressing in this manner, the observation at time-step 24 influences all states to time-step 24. These repeated updates are the source of the quadratic cost in current interval smoothing methods, that we propose a new method for.

Fixed lag smoothing seeks estimates only in a lag window (subfigure B). Thus the smoothed analysis at time step 0 utilizes all future observations in a window of W=8 model steps or L=2 observations. Equivalently, it can be said that an observation influences all states in a window  encompassed by W=8 prior model steps or L=2 prior observations. This is shown, for example, as observation at time step 20 influencing smoother analyses at time step 10.  The new algorithm proposed in this paper implement fixed-lag smoothing at a cost independent of the lag-length.          **46**

2.  Comparison of V1 and FBF fixed interval ensemble smoothing estimates with estimates from an ensemble Kalman filter (EnKF).  The error is computed between the smoothed ensemble analysis mean and truth. Observations are spaced every 5 model steps, the interval length is 100 and smoothed analyses are sought at every model time step. V1 and FBF give identical estimates; they only differ in computational requirements.  Smoothed estimates are consistently better than filter estimates inside the interval because they use all measurements rather than just those in the past…          **48**

3.  Comparison of V1-lag and FIFO-lag fixed lag ensemble smoothing estimates for different lag window lengths.  The other parameters are identical to those used in Figure 2. V1-lag and FIFO-lag give identical estimates; they only differ in computational requirements. Short fixed lags give results closer to the ensemble Kalman filter while longer fixed lags give results closer to the fixed interval smoother (compare to Figure 2)…          **49**

4. Computational times vs. fixed interval length. a) Model propagation only. b) Additional computational time (over model propagation) for ensemble Kalman filtering. c) Additional computational time (over model propagation and filtering) for V1 smoothing. d) Additional computational time (over model propagation and filtering) for FBF smoothing. The additional cost of V1 smoothing can be much more than filtering alone while the additional cost of FBF smoothing is minor. See text for detailed definition of each computational time… **50**

5. Computational times required to estimate states throughout fixed intervals of 100, 500, and 900 for FBF (fixed interval smoothing) and V1-lag (fixed lag smoothing). The FBF option (which does not depend on lag value) is shown at far left. V1-lag option is shown for a range of lags from 1 through 13. FBF is faster than V1-lag for lags greater than 2 ........ **51**

6. Computational times required to estimate states throughout fixed intervals of 100, 500, and 900 for FBF (fixed interval smoothing) and FIFO-lag (fixed lag smoothing). The FBF option (which does not depend on lag value) is shown at far left. The FIFO-lag option is shown for a range of lags from 1 through 13. FBF is faster than FIFO-lag. FIFO-lag computational time is nearly independent of lag (small fluctuations are related to random differences in time required to perform singular value decompositions at different lags)… **52**

**(A) Graphical Model of Fixed-Interval Smoothing**

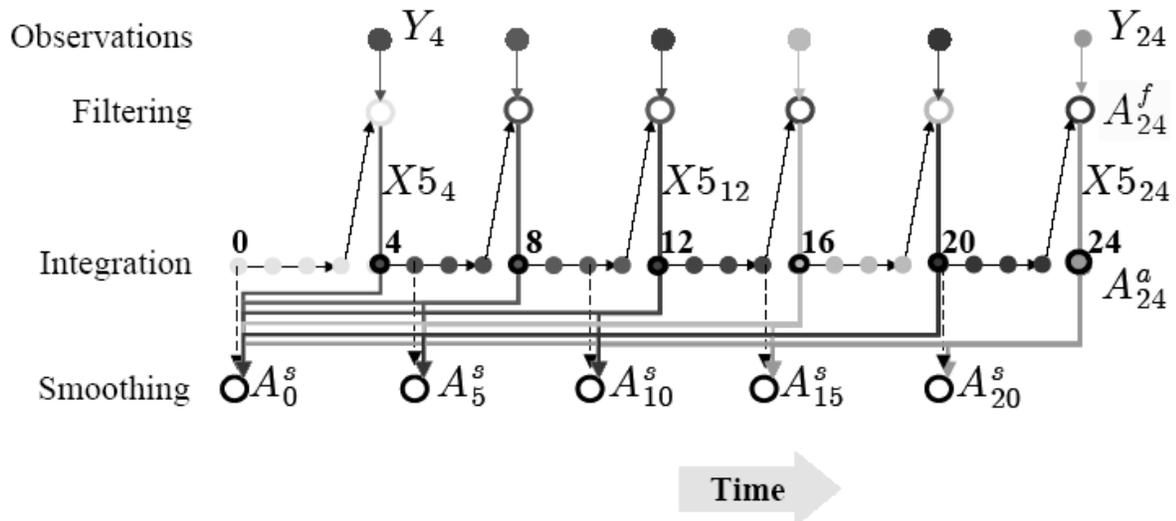

**(B) Graphical Model of Fixed-Lag Smoothing**

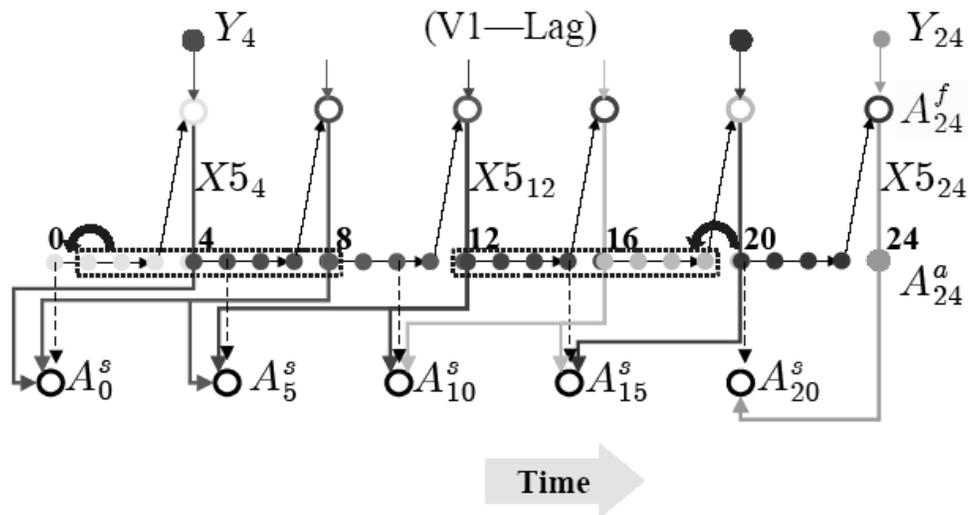

**Figure 1:** Illustration of smoothing problems. Subfigure (A) depicts fixed-interval smoothing and subfigure (B) depicts fixed-lag smoothing. In subfigure (A), there are M=24 model integrations, with R=6 observations at r=4 equally spaced model steps 4,8,12,16,20 and 24. This figure is a graphical model for interval smoothing. It shows that the ensemble is integrated

forward starting at time-step 0 (see "Integration"). When an observation is encountered (see row "Observations"), filtering is performed. For example, at $Y_4$ the corresponding $X5_4$ is generated and used to update the forecast ensemble ( $A_4^f$ ). It is also used to update forecast (or filtered) ensembles at all previous times where smoothed analyses are desired. In this example smoothed analyses are desired at time steps 0, 5, 10, 15 and 20 and may be between or at measurement times. Using the depicted scheme, it can be seen that the smoothed analysis at time 0 utilizes all 6 future observations, the smoothed analysis at time step 10 uses four future observations (filtering already incorporates all observations to time step 10), and so on. Equivalently, the observation at time step 4 influences all states at time steps 0-4 and, progressing in this manner, the observation at time-step 24 influences all states to time-step 24. These repeated updates are the source of the quadratic cost in current interval smoothing methods, that we propose a new method for.

Fixed lag smoothing seeks estimates only in a lag window (subfigure B). Thus the smoothed analysis at time step 0 utilizes all future observations in a window of W=8 model steps or L=2 observations. Equivalently, it can be said that an observation influences all states in a window encompassed by W=8 prior model steps or L=2 prior observations. This is shown, for example, as observation at time step 20 influencing smoother analyses at time step 10. The new algorithm proposed in this paper implement fixed-lag smoothing at a cost independent of the lag-length.

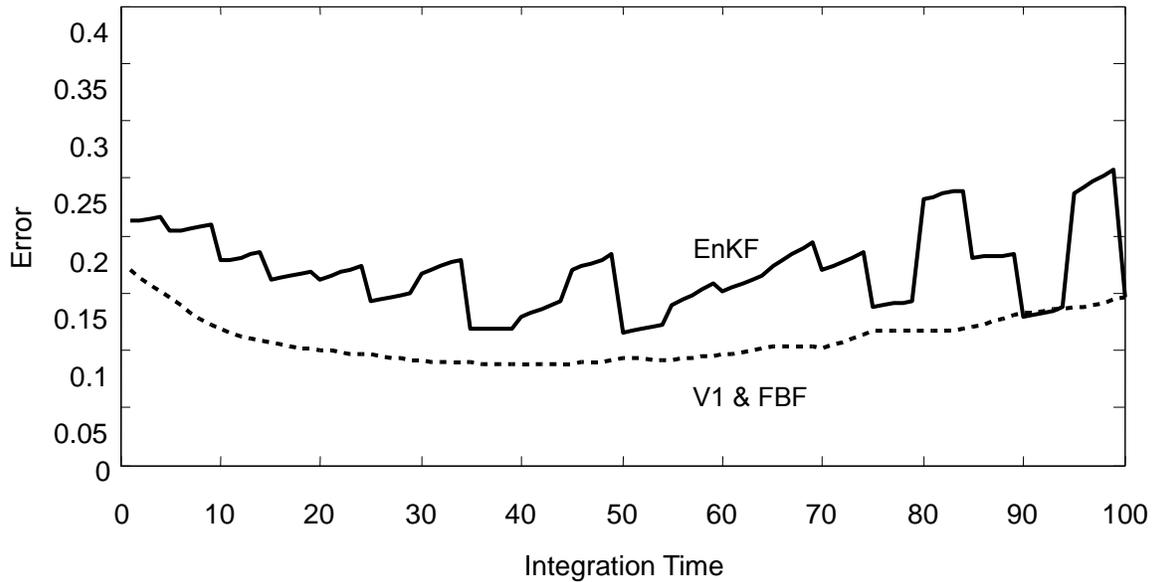

**Figure 2:** Comparison of V1 and FBF fixed interval ensemble smoothing estimates with estimates from an ensemble Kalman filter (EnKF). The error is computed between the smoothed ensemble analysis mean and truth. Observations are spaced every 5 model steps, the interval length is 100 and smoothed analyses are sought at every model time step. V1 and FBF give identical estimates; they only differ in computational requirements. Smoothed estimates are consistently better than filter estimates inside the interval because they use all measurements rather than just those in the past.

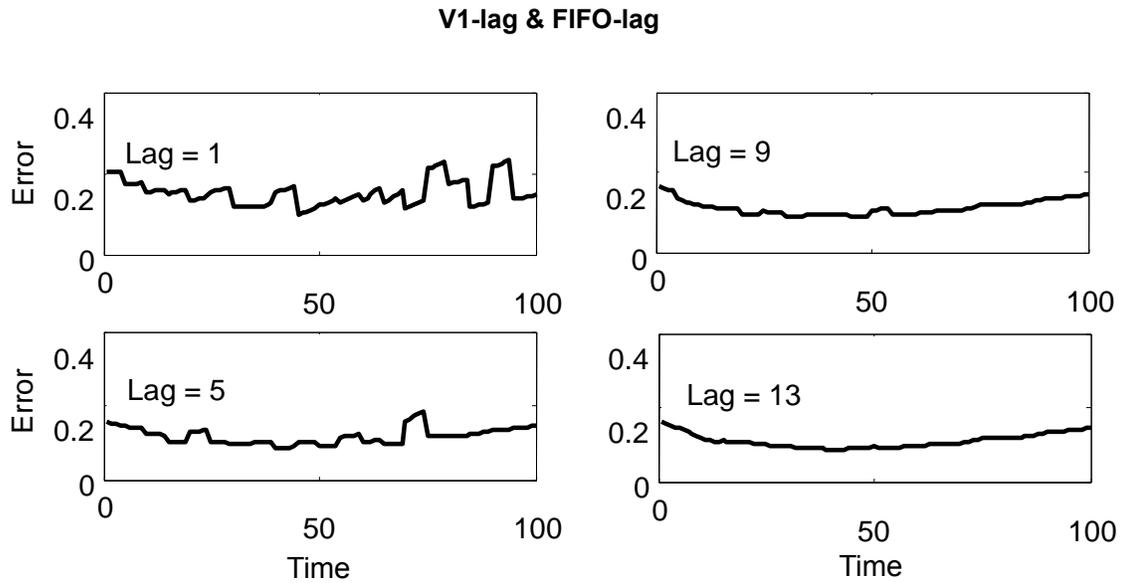

**Figure 3:** Comparison of V1-lag and FIFO-lag fixed lag ensemble smoothing estimates for different lag window lengths. The other parameters are identical to those used in Figure 2. V1-lag and FIFO-lag give identical estimates; they only differ in computational requirements. Short fixed lags give results closer to the ensemble Kalman filter while longer fixed lags give results closer to the fixed interval smoother (compare to Figure 2).

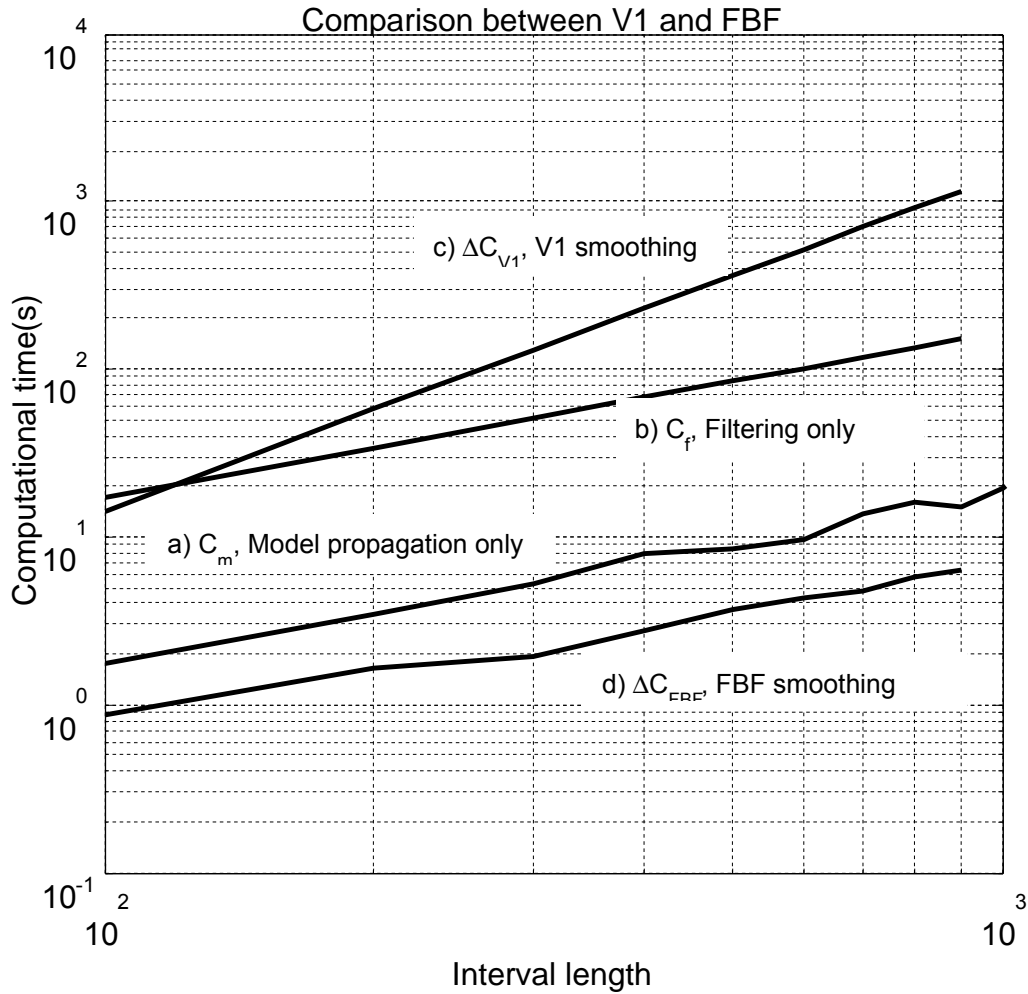

**Figure 4:** Computational times vs. fixed interval length. a) Model propagation only. b) Additional computational time (over model propagation) for ensemble Kalman filtering. c) Additional computational time (over model propagation and filtering) for V1 smoothing. d) Additional computational time (over model propagation and filtering) for FBF smoothing. The additional cost of V1 smoothing can be much more than filtering alone while the additional cost of FBF smoothing is minor. See text for detailed definition of each computational time.

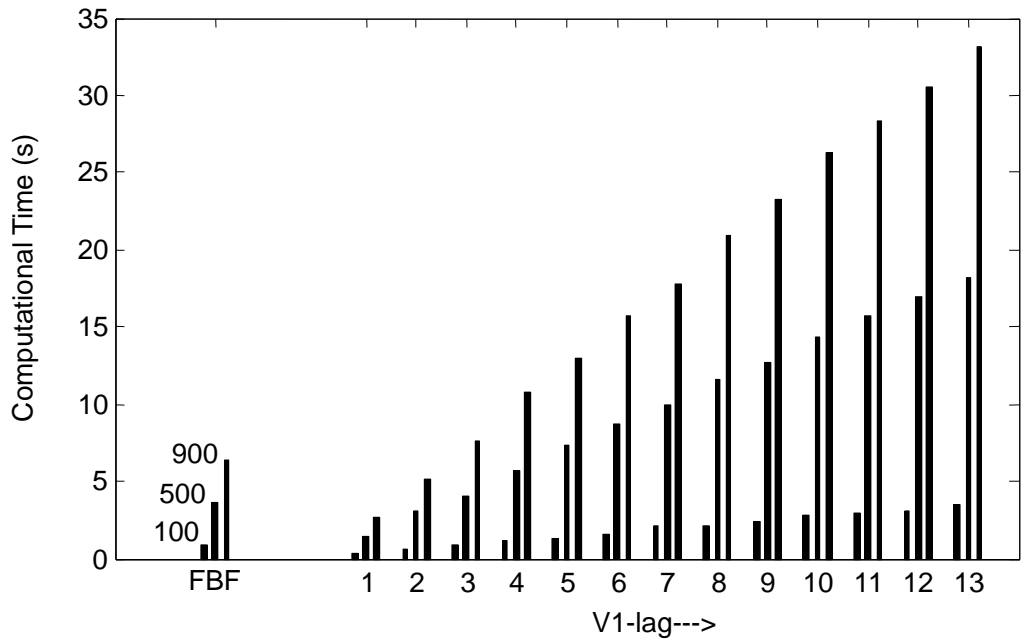

**Figure 5:** Computational times required to estimate states throughout fixed intervals of 100, 500, and 900 for FBF (fixed interval smoothing) and V1-lag (fixed lag smoothing). The FBF option (which does not depend on lag value) is shown at far left. V1-lag option is shown for a range of lags from 1 through 13. FBF is faster than V1-lag for lags greater than 2.

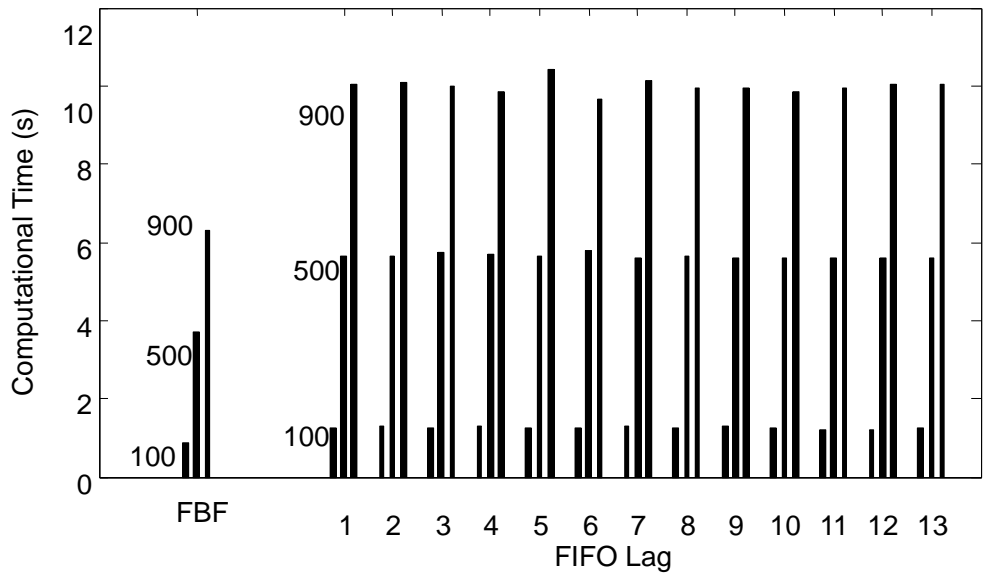

**Figure 6:** Computational times required to estimate states throughout fixed intervals of 100, 500, and 900 for FBF (fixed interval smoothing) and FIFO-lag (fixed lag smoothing). The FBF option (which does not depend on lag value) is shown at far left. The FIFO-lag option is shown for a range of lags from 1 through 13. FBF is faster than FIFO-lag. FIFO-lag computational time is nearly independent of lag (small fluctuations are related to random differences in time required to perform singular value decompositions at different lags).